\documentclass[12pt,preprint,useAMS,usenatbib]{emulateapj}
\usepackage{graphicx}
\usepackage{subfloat}
\usepackage{amsmath}
\def\msun{{M_\odot}}
\def\mdot{\,{M_\odot\,{\rm yr^{-1}}}}

\def\spose#1{\hbox to 0pt{#1\hss}}
\def\lta{\mathrel{\spose{\lower 3pt\hbox{$\mathchar"218$}}
     \raise 2.0pt\hbox{$\mathchar"13C$}}}
\def\gta{\mathrel{\spose{\lower 3pt\hbox{$\mathchar"218$}}
     \raise 2.0pt\hbox{$\mathchar"13E$}}}

\newcommand{\etal}{{et al.\ }}

\def\cm3{\,{\rm cm^{-3}}}

\shortauthors{Madau, Weisz, and Conroy}
\submitted{ApJL, in press}


\begin{document}

\title{Reversal of Fortune: Increased Star Formation Efficiencies in the Early Histories of Dwarf Galaxies?}
\author{Piero Madau$^1$, Daniel R. Weisz$^{1,2,3}$, and Charlie Conroy$^1$} 
\altaffiltext{1}{Department of Astronomy and Astrophysics, University of California, 1156 High Street, Santa Cruz, CA 95064, USA.}
\altaffiltext{2}{Department of Astronomy, University of Washington, Box 351580, Seattle, WA 98195, USA.}
\altaffiltext{3}{Hubble Fellow.}

\begin{abstract}
On dwarf galaxy scales, the different shapes of the galaxy stellar mass function and the dark halo mass function require a star-formation efficiency (SFE)
in these systems that is currently more than 1 dex lower than that of Milky Way-size halos. Here, we argue that this trend may actually be reversed at 
high redshift. Specifically, by combining the resolved star-formation histories of nearby isolated dwarfs with the simulated mass-growth rates of 
dark matter halos, we show that the assembly of these systems occurs in two phases: (1) an early, fast halo accretion phase with a rapidly deepening 
potential well, characterized by a high SFE; and (2) a late slow halo accretion phase where, perhaps as a consequence of reionization, the SFE is low.
Nearby dwarfs have more old stars than predicted by assuming a constant or decreasing SFE with redshift, a behavior that appears to deviate 
qualitatively from the trends seen amongst more massive systems. Taken at face value, the data suggest that, at sufficiently early epochs, 
dwarf galaxy halos above the atomic cooling mass limit can be among the most efficient sites of star formation in the universe. 
\end{abstract}

\section{Introduction}\label{intro}

Dwarf galaxies are the smallest, most abundant, least luminous systems in the universe and have come to play a critical role in our understanding 
of the mapping from dark matter halos to their baryonic components \citep{Pontzen14}. An abundance mismatch problem has emerged over the past two decades: 
the different shapes of the galaxy stellar mass function and the dark halo mass function on dwarf galaxy scales require a star-formation efficiency (SFE)
of only $\sim 0.1$\% in these systems \citep[e.g.,][]{Behroozi13}, a value that has been traditionally difficult to reproduce in hydrodynamical 
simulations \citep[e.g.,][]{Sawala11}. A strongly decreasing stellar mass fraction with decreasing halo mass is also required to solve the ``missing satellite 
problem", the discrepancy between the small number of dwarf satellites orbiting the Milky Way and the vastly larger number of dark 
matter subhalos predicted to survive in $\Lambda$CDM \citep[e.g.,][]{Moore99,Klypin99,Diemand08,Koposov09,Rashkov12}.

Many astrophysical processes that suppress gas accretion and star formation in dwarf galaxies have been proposed to solve this puzzle, including the heating of 
intergalactic gas by the ultraviolet photoionizing background \citep[e.g.,][]{Efstathiou92,Bullock00,Kravtsov04}, rapid mass loss driven by supernovae 
(SNe) \citep[e.g.,][]{Dekel86,Mori02}, and metallicity-dependent H$_2$-regulation \citep[e.g.,][]{Gnedin10,Kuhlen13}. Despite the successes 
of a new generation of hydrodynamical simulations in bringing theoretical predictions in better agreement with many observations 
\citep[e.g.,][]{Mashchenko08,Governato10,Shen14,Madau14}, a characterization of the time-dependent role of these mechanisms remains elusive. 
In particular, to reconcile the low SFEs of nearby dwarfs with the steep faint-end slope of the galaxy ultraviolet luminosity 
function measured at $z\gta 5$ \citep{Bouwens12} and observational constraints on reionization \citep{Hinshaw13,Robertson13}, the star formation rate in low-mass halos 
must be enhanced  at early times relative to that at lower redshifts \citep{Lu14}.  

\begin{figure}
\centering
\includegraphics[width=0.49\textwidth]{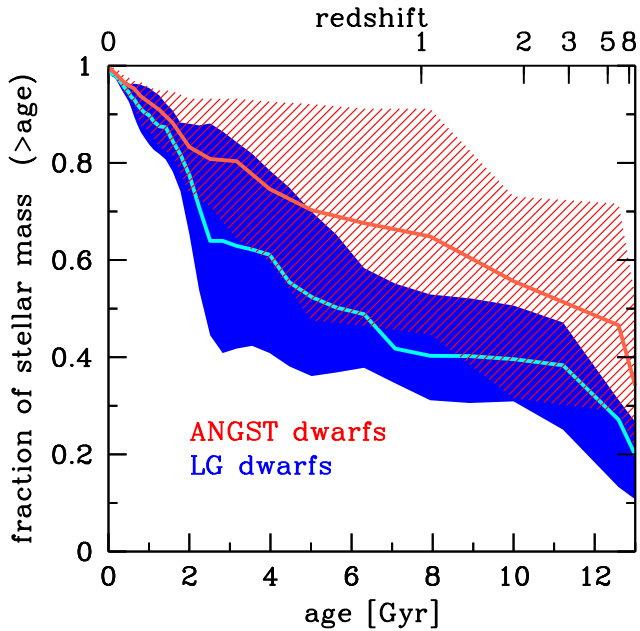}
\vspace{0.0cm}
\caption{Cumulative stellar mass fraction as a function of lookback time (bottom axis) and redshift (top axis).
{Blue colors}: Isolated dwarfs from the Local Group sample of \citet{Weisz14}.
The solid line is the mean of the distribution of best-fit star-formation histories for individual galaxies and the colored envelope 
marks the 16th and 84th percentiles, i.e., a (statistical) 68\% confidence interval around the mean. 
{Red colors:} Same for dwarf galaxies in the ANGST survey \citep{Weisz11}.
}
\label{fig1}
\end{figure}

In recent years, large systematic surveys of dwarfs in and around the Local Group (LG) with the {\it Hubble Space Telescope (HST)}, such as the ACS Nearby Galaxy Survey 
Treasury (ANGST) \citep{Dalcanton09}, have uniformly measured star-formation histories (SFHs) for over 100 low-mass systems \citep{Weisz11,Weisz14}. In this Letter 
we use results from these data sets to investigate further the interplay between the dark matter and stellar assembly histories of these systems.

\section{Stellar and Dark Matter Assembly Histories}

We consider here the SFHs of nearby dwarf galaxies, $M_\star(z=0)\lta 10^8\,\msun$, located $>300$ kpc from a massive host (such as the Milky Way). These isolated 
systems  provide a template for how low-mass galaxies evolve in the absence of significant environmental influence.\footnote{Nearby 
isolated dwarfs are mostly dwarf irregulars by morphological type.  Although dwarf irregulars are gas-rich and have ongoing star formation, most evidence suggests that 
their properties are similar to dwarfs of other morphological classes, such as dwarf spheroidals, and have just not been environmentally processed by a massive 
host \citep[e.g.,][]{Weisz11,Weisz14,Kirby14}.}\, Figure \ref{fig1} depicts the cumulative SFHs, i.e., the fraction of total stellar mass formed prior to a given 
epoch, of dwarfs in the LG and nearby field (ANGST) \citep{Weisz11,Weisz14}. The solid line represents the {\it mean} of the distribution of best-fit SFHs
and the error envelope marks the 68\% confidence interval of the mean. 

On average, LG dwarfs formed $\gta$ 30\% of their stellar mass prior to redshift 2 ($>$10--11 Gyr ago) and show increasing rate of stellar mass growth toward the 
present beginning around redshift 1. However, as emphasized in \citet{Weisz14}, the {\it HST}-based SFHs of many LG dwarfs likely only provide \emph{lower limits} on 
star formation at early times owing to observational aperture effects.  To maximize the number of stars for SFH measurements, {\it HST} observations have typically targeted the 
central, high surface brightness regions of dwarfs and, owing to the limited size of the {\it HST} field-of-view, exclude most of their ancient stellar 
halos \citep[e.g.,][]{Hidalgo13}. This effect is clearly seen in the different mean ANGST and LG SFHs: observations of ANGST galaxies cover a larger fraction of the 
system and include more old (halo) stars relative to LG dwarfs. 
An increase in the observational aperture would lead to an \emph{increase} in the relative amount of stellar mass formed prior to 10 Gyr ago in many isolated dwarfs.

Despite the improved areal coverage of ANGST dwarfs, uncertainties on their SFHs are a factor of $\sim$ 2 larger than for LG dwarfs for ages $>$ 10 Gyr. The 
increased uncertainties are due to the relatively shallow color-magnitude diagrams (CMDs) of ANGST systems, which result in larger systematic uncertainties 
(i.e., the variation in SFH when measured with different stellar libraries) on the fraction of old stars. The characteristic ANGST CMD partially resolves the red clump, 
meaning the SFHs for ages older than a few Gyr rely on evolved phases of stellar evolution (e.g., red giant branch) where systematic uncertainties are larger.  

In comparison, LG CMDs typically extend several magnitudes deeper than ANGST, 
providing access to the horizontal branch, which is sensitive to star formation $>$ 10 Gyr ago, and to older and better-understood main-sequence stars, thereby 
reducing the systematic uncertainties at older ages.  The best possible constraints on the ancient SFHs \citep[$\lta$ 1 Gyr age resolution at all ages; 
e.g.,][]{Gallart2005} come from CMDs that include the oldest main sequence turnoff (MSTO).  However, owing to the faintness of this feature 
($M_V\sim +4$) and the effects of stellar crowding, it is challenging to observe the oldest MSTO outside of the closest satellite galaxies, even with the {\it HST}. 
As a result, only a handful of isolated dwarfs have maximally secure SFHs.  We discuss some of these systems in Section \ref{sec:SFE}.

\begin{figure*}
\centering
\includegraphics[width=0.49\textwidth]{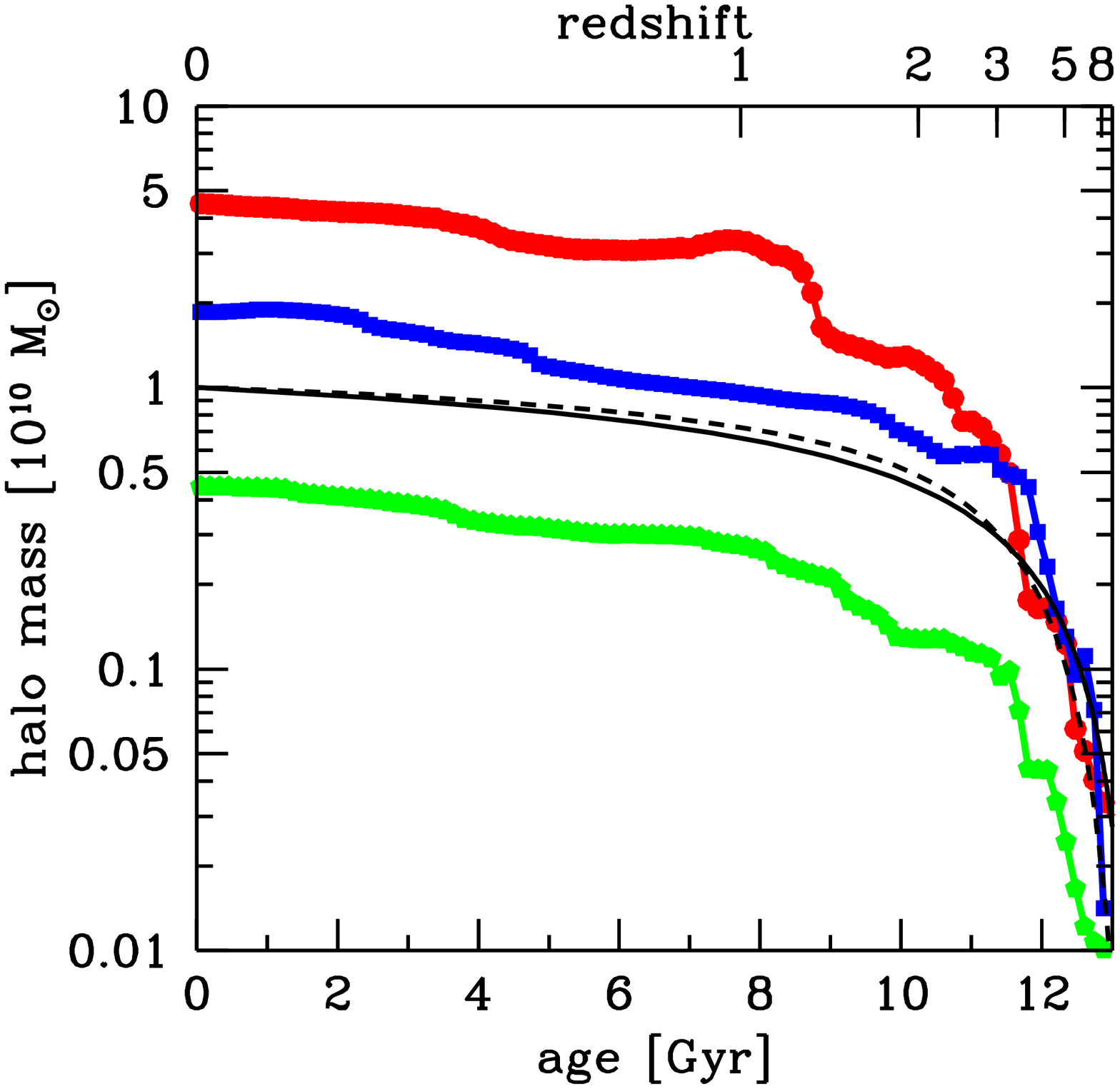}
\includegraphics[width=0.49\textwidth]{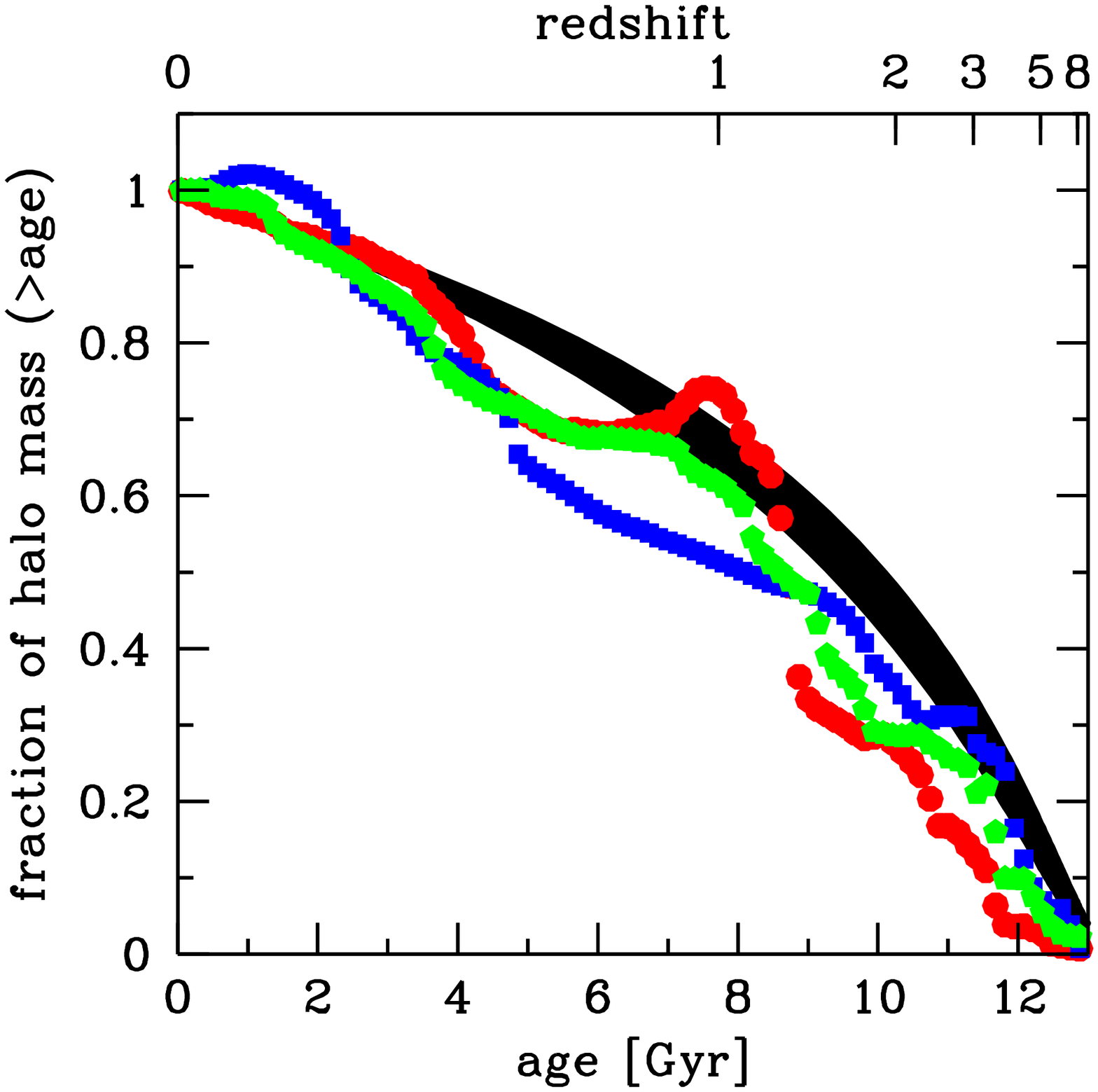}
\vspace{0.0cm}
\caption{Mass-assembly history of dwarf dark matter halos. {\it Left panel:} Halo mass $M(z)$ as a function of lookback time. {\it Colored points:} Assembly histories for 
three massive dwarfs (Bashful, Doc, and Dopey) simulated by \citet{Madau14} (DM-only run)
at a particle mass resolution of $m_{\rm DM}=1.9\times 10^4\,\msun$. The simulation was performed with the parallel TreeSPH code \textsc{Gasoline} \citep{Wadsley04}.
{\it Solid black curve}: Mean mass-assembly history, $\langle M(z)\rangle$ for all $z=0$ $M_0=10^{10}\,\msun$ halos in the two Millennium simulations \citep{Fakhouri10}. 
{\it Dashed black curve}: Mean mass-assembly history for $M_0=10^{10}\,\msun$ halos obtained using the best-fit mass growth rates of \citet{Behroozi14}, which 
are based on the {\it Bolshoi} simulation \citep{Klypin11}.
{\it Right panel:} Cumulative halo mass fraction as a function of lookback time. {\it Colored points}: Same as left panel.   
The dark swath bounds the mean mass-assembly histories for all $M_0=3\times 10^9\,\msun$ ({\it top curve}) and $M_0=3\times 10^{10}\,\msun$ ({\it bottom curve}) 
halos in the Millennium simulations. 
}
\label{fig2}
\end{figure*}

To empirically connect the buildup of stellar mass in dwarfs to the assembly of their host dark matter halos over cosmic time predicted by $\Lambda$CDM, we rely here on the 
merger trees and growth rates derived by \citet{Fakhouri10} and based on the joint data set from the Millennium and Millennium-II simulations. The best-fit to the {\it mean} 
dark matter accretion rate of halos of mass $M(z)$ at redshift $z$ is
\begin{equation}
\langle \dot M\rangle=(46.1\mdot)\,M_{12}^{1.1}\,(1+1.11z)\,{H(z)\over H_0}, 
\end{equation} 
where $M_{\rm 12}\equiv M/(10^{12} \msun)$ and $H(z)$ is the Hubble parameter, $H(z)=H_0\sqrt{\Omega_M(1+z)^3+\Omega_\Lambda}$. Figure \ref{fig2} shows 
the mean mass-assembly history $\langle M(z)\rangle$, obtained by integrating the above fitting formula in a $(\Omega_M,\Omega_\Lambda)=(0.3,0.7)$ cosmology, of 
all $z=0$ $M_0=10^{10}\,\msun$ halos in the two Millennium simulations. \citet{Fakhouri10} assigned masses using the standard friends-of-friends algorithm 
and their growth rates must be extrapolated beyond $z=6$ because of numerical resolution effects. We have compared this assembly history with that derived for
the same $M_0$ using an alternative mass definition (based on spherical overdensity) and the best-fit mass growth rates of \citet{Behroozi14}, and find 
consistent results. 

As a further consistency check, we have also plotted the (spherical overdensity) mass-assembly histories of three $3\times 
10^9\,\msun\lta M_0\lta 4.5\times 10^{10}\,\msun$ individual dwarf halos simulated at much higher mass resolution (approximately 500 times better than Millennium-II) in a 
fully cosmological setting \citep{Madau14}. {All dwarf halos grow quickly at early times, by approximately one order of magnitude in less than 1 Gyr}. This is a 
consequence of the shallower slope of the power spectrum on these mass scales: dark matter clumps of all masses collapse nearly simultaneously and the timescale 
between collapse and subsequent merging becomes shorter. The early rapid dark matter accretion phase is followed, at redshift $z\lta 3$, by a phase of 
slower growth characterized by the gentle addition of mass onto an established potential well.\footnote{``Spherical overdensity masses" are known to undergo 
spurious ``pseudo-evolution" owing to the changing reference density \citep{Diemand07}. \citet{Diemer13} have recently shown that this pseudo-evolution 
accounts for almost the entire mass growth of galaxy-size halos between $z=1$ and today.}\,
In Figure \ref{fig2} we also show the cumulative mass fraction of the simulated dwarfs as a function of lookback time. The dark band 
provides an indication of the variation in the mean assembly histories of all $3\times 10^9\,\msun\le M_0\le 3\times 10^{10}\,\msun$ halos in the Millennium 
simulations \citep{Fakhouri10}.

\begin{figure*}
\centering
\includegraphics[width=0.49\textwidth]{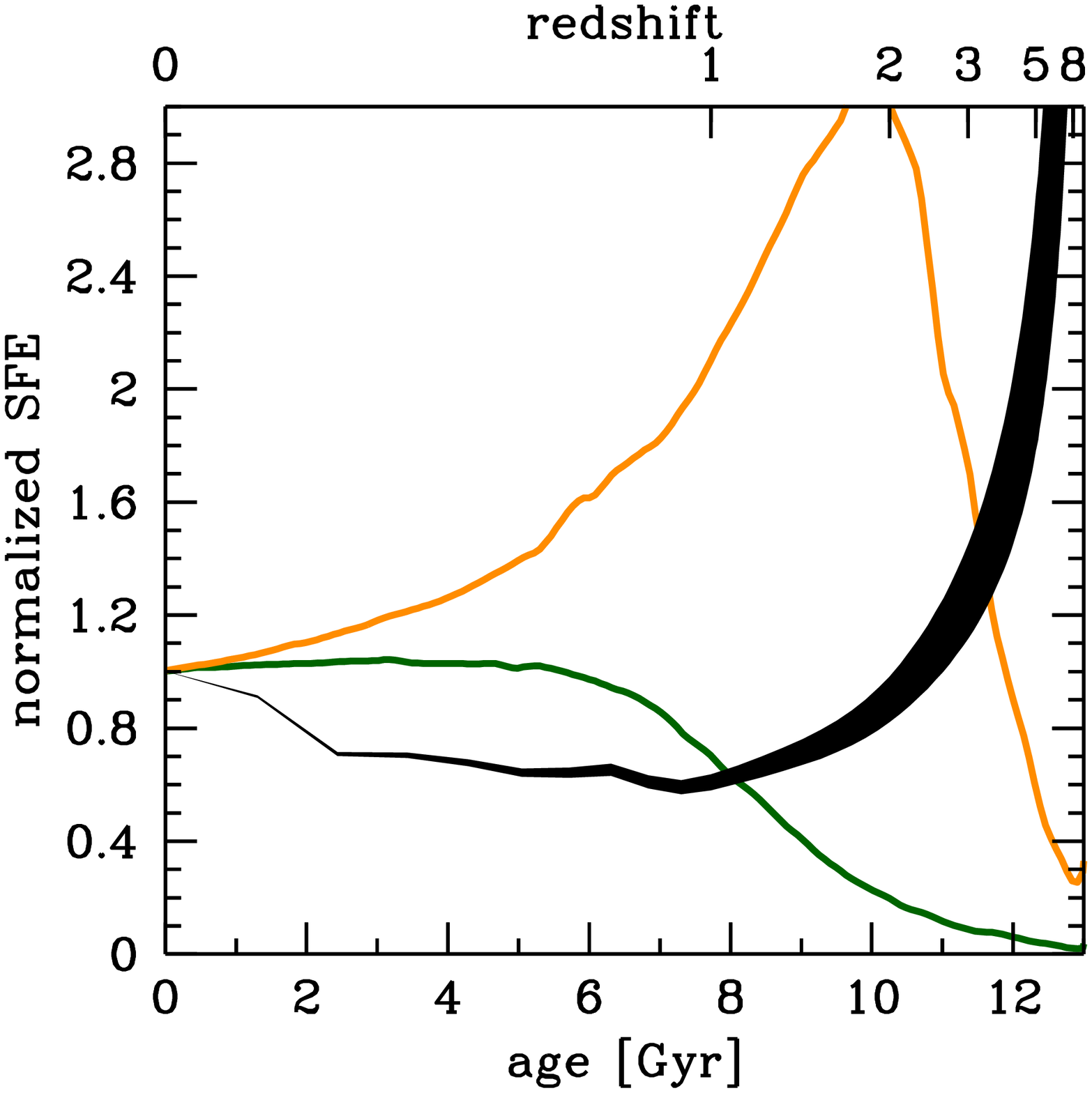}
\includegraphics[width=0.49\textwidth]{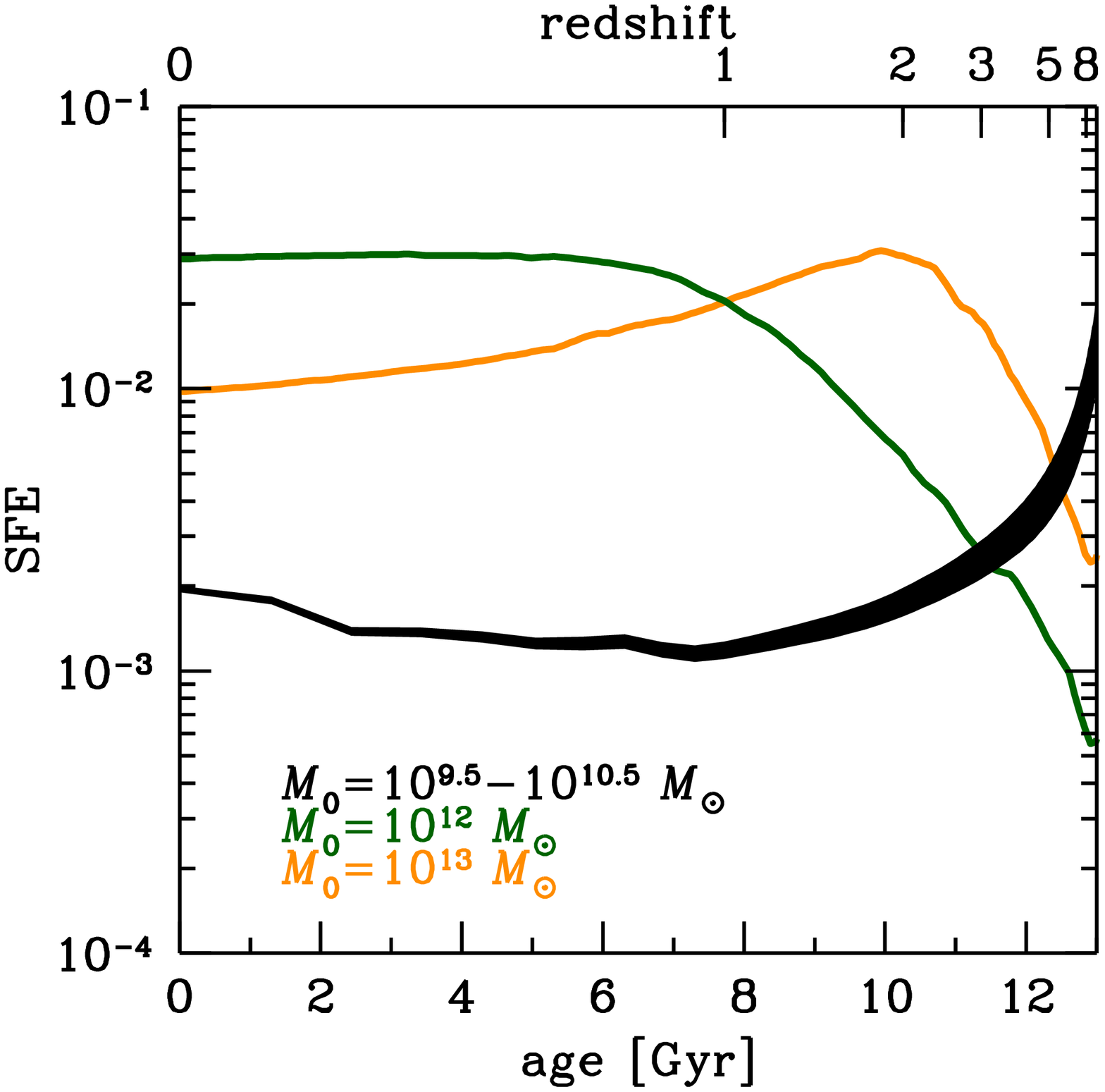}
\vspace{0.0cm}
\caption{{Left panel}: Mean SFEs of galaxies, normalized to their today's values. {Dark band}: Isolated Local Group dwarf galaxies. The strip shows the ratio 
estimator obtained by dividing the mean SFH of dwarfs by the mean mass-assembly history of halos with $3\times 10^9\,\msun\le M_0\le 3\times 10^{10}\,\msun$ in 
the Millennium simulations. The green and orange bands show the SFE versus lookback time of a Milky Way-size halo, $M_0=10^{12}\,\msun$, and of an elliptical 
galaxy halo, $M_0=10^{13}\,\msun$, inferred by the abundance matching technique of \citet{Behroozi13}. {Right panel}: Same as the left panel, not normalized. 
The present-day efficiency of dwarf galaxies was fixed to $2\times 10^{-3}$, the value inferred for $M_0\simeq 10^{10}\,\msun$ dwarfs by \citet{Behroozi13}.  
}
\label{fig3}
\end{figure*}

A look at Figures \ref{fig1} and \ref{fig2} reveals that the steep drop in halo mass at lookback times $>10$ Gyr is not accompanied by an equally sharp drop in stellar mass. 
Conversely, the increase in the stellar content of LG dwarfs that is inferred at these early epochs is far more modest that the predicted, more than tenfold 
increase in halo mass with cosmic time. {This is followed, at ages between 7 and 11 Gyr, by a plateau in the SFH where the star formation activity actually 
declines.} As a consequence and as detailed below, the SFE ($\equiv M_*/M$) of dwarf galaxies decreases rapidly from high redshift to the present day.

\section{Star Formation Efficiencies} 
\label{sec:SFE}  

To illustrate this point quantitatively and shed light on the dwarf galaxy-halo connection at different epochs, we have plotted in Figure \ref{fig3} the mean stellar 
mass per halo mass for dwarf galaxies (i.e. the ratio estimator obtained by dividing the mean SFH of isolated LG dwarfs by the mean mass-assembly history of halos with 
a given $M_0$), as a function of lookback time. We assume that dwarfs occupy host halos in the mass range $3\times 10^9\,\msun\le M_0\le 3\times 10^{10}\,\msun$; as 
shown in Figure \ref{fig2}, such halos have very similar mass assembly histories.
It is clear that dwarfs go through markedly different phases of galaxy growth. Specifically, the early progenitor halos of today's dwarfs were relatively more 
efficient at converting gas into stars. The SFE drops with cosmic time from an unresolved peak at high redshift, as dwarfs form stars less rapidly (or not at 
all at ages between 7 and 11 Gyr) than their host halos accrete dark matter. Below $z\sim2$, the SFE remains approximately constant, as the increase in stellar 
mass approximately tracks the growth of the host dark matter halo.\footnote{Note that we implicitly assume that all star formation is in-situ, i.e., 
took place in the main host. In principle, a significant fraction of stars may have formed in progenitor halos and have been accreted through merging. 
In practice, this effect will be small because the SFE is a steeply declining function of halo mass and so accreted halos will bring in very few stars.}

A decreasing SFE from an early peak paints a different picture of galaxy growth than envisioned for more massive galaxies. Figure \ref{fig3} also shows, for comparison, 
the SFE versus lookback time inferred by the version of the ``abundance matching" technique of \citet{Behroozi13} for a Milky Way-size halo, $M_0=10^{12}\,\msun$, 
and for a massive 
galaxy halo, $M_0=10^{13}\,\msun$. The former is characterized by a late plateau, extending for $\sim$ 6 Gyr, and by a monotonic decline for $z\gta 1$. The latter reaches 
a maximum at $z\approx 2$ and drops dramatically at higher redshift. The progenitors of these massive halos had early SFEs that were much lower than their present-day 
values, i.e. they were forming stars inefficiently prior to 11 Gyr ago. By contrast, at these early epochs, the progenitors of LG dwarfs appear to have followed a divergent 
path, forming stars with a SFE that was significantly higher than today. These findings clearly demonstrate the dangers of trying to infer the properties of low-mass galaxies 
from their higher mass counterparts. 
Figure \ref{fig3} provides another perspective to the tension between the CMD-based SFHs of nearby dwarfs and the spectra-based SFHs of more massive star-forming 
Sloan Digital Sky Survey (SDSS) galaxies highlighted by \citet{Leitner12}. Resolved dwarfs appear to assemble their stellar content earlier than more massive 
systems, which form late and had only $\sim 15\%$ of their stellar mass in place before $z=1-2$.  Taking the data at face value, dwarfs galaxies clearly do not lie at one end 
of a continuum defined by the more massive galaxies. Instead, they seem to show a behavior that deviates qualitatively from the trends seen on larger scales. 
Qualitatively similar conclusions have been reached recently by \citet{Behroozi14}: based on galaxies' 
specific star-formation rates, halos smaller than $10^{12}\,\msun$ are predicted in this study to form stars with increasing efficiencies at $z>4$.

\begin{figure*}
\centering
\includegraphics[width=0.49\textwidth]{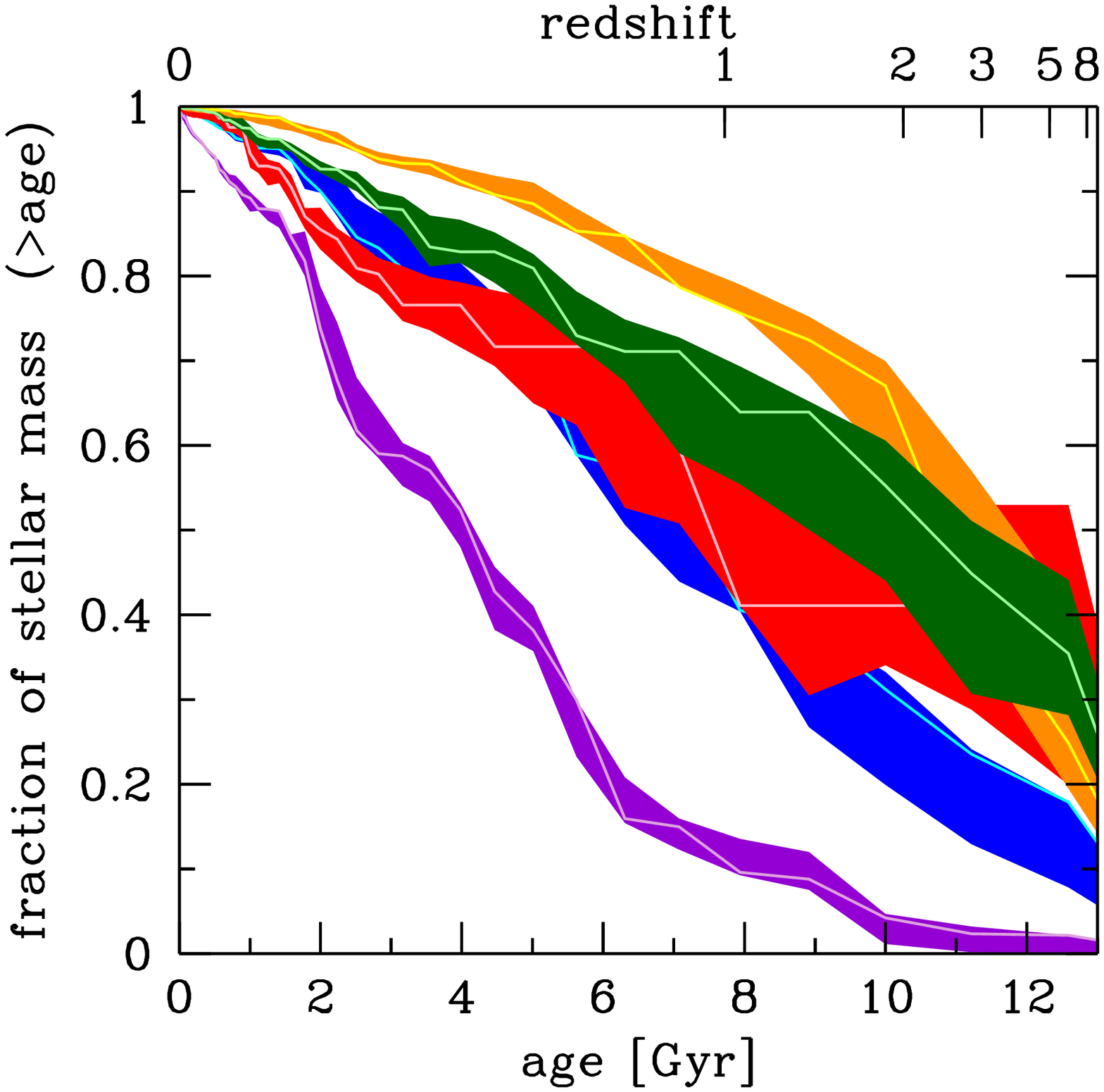}
\includegraphics[width=0.49\textwidth]{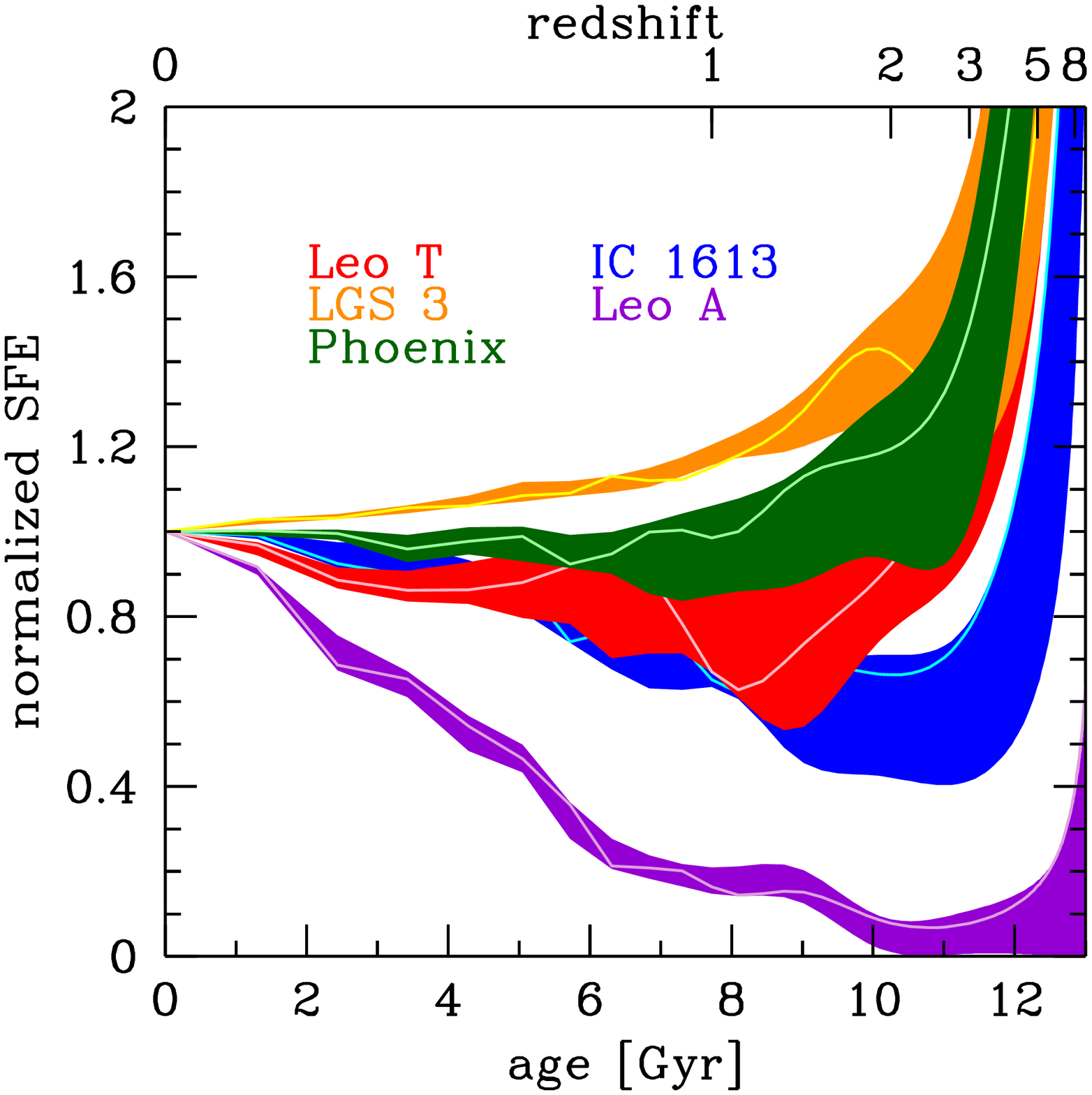}
\vspace{0.0cm}
\caption{Cumulative stellar mass fraction ({left panel}) and SFE ({right panel}) as function of lookback time (bottom axis) and redshift (top axis) for 
five gas-rich LG dwarfs with CMDs that extend below the oldest MSTO. The colored envelopes mark the 68\% confidence interval around the best-fit SFH, including systematic 
uncertainties. The SFEs in the right panel assume the mean mass-assembly history of a $M_0=10^{10}\,\msun$ dwarf halo. The SFHs of Leo~A, Phoenix, LGS~3, and IC~1613 
were originally derived as part of the Local Cosmology for Isolated Dwarfs programs \citep{Cole07,Hidalgo09,Hidalgo11,Skillman14}.  However, to ensure self-consistency 
with the other LG and ANGST SFHs, we use here the SFHs of these galaxies that were re-measured by \citet{Weisz14}.
}
\label{fig4}
\end{figure*}

To illustrate the robustness of our results, we have compared the SFEs obtained from the average SFHs of LG dwarfs with those derived from deeper {\it HST}/ACS observations, 
i.e. from CMDs that extend below the oldest MSTO. In Figure \ref{fig4} we consider five such isolated dwarfs whose SFHs have minimal uncertainties (less than 10\%
of their lookback ages): IC 1613, Leo A, Leo T, 
LGS 3, and Phoenix. Overall, we find that the deep ACS-based SFEs are similar to those derived from shallower data: all but Leo A show the same trend, an SFE that remains 
constant or decreases from the present to redshift 1.5 and then grows rapidly prior to 10 Gyr ago. 

\section{Discussion}   

At present, dwarf galaxies are among the least efficient star-forming systems in the universe. And yet, they have more old stars than predicted by assuming a constant or 
decreasing SFE with increasing redshift, i.e., if they followed the same trend inferred for Milky Way-size halos by the abundance matching technique. 
Although there is some scatter in the fraction of ancient stars (with known exceptions to these trends such as Leo A), the typical isolated LG dwarf appears to have formed one 
fourth of its present-day stellar content more than 12.5 Gyr ago. This high fraction of stellar mass at earlier 
times is in contrast to results derived for more massive galaxies and suggests that nearby dwarfs do not follow the clear ``downsizing" trend whereby more massive 
galaxies form the bulk of their stars at earlier epochs compared to lower mass galaxies \citep{Leitner12,Weisz14}.  The physics behind such a different behavior remains 
to be understood, but the implications are clear: the early progenitors of today's dwarfs had a higher stellar content per unit dark mass than their low-redshift 
counterparts of similar size. This cannot be true below virial masses of $10^8\,\msun$ or so: such ``mini-halos" must have formed stars 
very inefficiently not to overproduce the abundance of ultra-faint dwarf satellites of the Milky Way \citep{Madau08}. Taken together, these 
trends suggest that, at sufficiently high redshift, dwarf galaxy halos above the atomic cooling limit (i.e., massive enough to cool via collisional 
excitation of hydrogen Lyman-$\alpha$) can be among the most efficient sites of star formation in the universe. 

\begin{figure}
\centering
\includegraphics[width=0.49\textwidth]{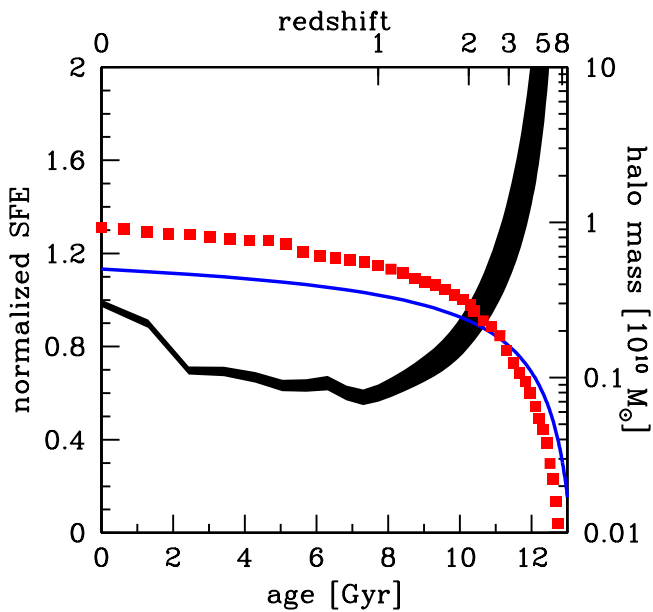}
\vspace{0.0cm}
\caption{Comparison between the mean normalized SFE of LG dwarfs ({dark band}, units on the left axis) and the evolution of the characteristic mass 
$M_c$ ({red points}, units on the right axis) below which dwarf galaxy halos lose half of their gas from photoheating \citep{Okamoto08}.
The blue curve shows the mean mass-assembly history of a $M_0=5\times 10^{9}\,\msun$ dwarf halo. 
}
\label{fig5}
\end{figure}

The assembly of dwarf galaxy systems appears then to occur in two main phases: (1) an early, fast halo accretion phase with a rapidly deepening potential
well, characterized by a high SFE; and (2) a late slow halo accretion phase when the SFE is low. The early peak in SFE is not resolved by current observations: 
SFHs with a sub-Gyr age resolution should show a maximum and a drop at very early times as the masses of dwarf progenitors fall 
below the atomic cooling limit.     

A useful perspective on these findings may be obtained by looking at mechanisms that modulate the supply of cold gas and limit the formation of 
stars in dwarf galaxies. Photoionization by the cosmic UV background, for example, heats the intergalactic medium to temperatures $T\gta 10^4\,$K
and reduces line cooling rates by lowering the fraction of neutral atoms. Both effects influence the ability of gas 
to accrete, cool, and condense in low-mass systems and can set a minimum mass scale for galaxy formation \citep{Efstathiou92}. The critical halo size below 
which such suppression is important has been addressed by several authors \citep[e.g.,][]{Thoul96,Kitayama00,Gnedin00,Dijkstra04}.
Cosmological hydrodynamical simulations by \citet{Okamoto08} have shown that the characteristic mass, $M_c(z)$, below which dwarf galaxy halos lose half of their gas from 
photoheating rises from $\sim 1.5\times 10^7\,\msun$ after reionization to $\sim 10^{10}\,\msun$ at $z=0$. In Figure \ref{fig5} we plot this characteristic mass together 
with the mean mass-assembly history of a $M_0=5\times 10^9\,\msun$ halo. The comparison shows how a substantial amount of gas may be able to cool and form stars at early
times, when $M>M_c$. Star formation may be sustained at high SFE for a couple of Gyr until the halo gas content is severely depleted ($M_c>M$), leading to a decline in the 
star formation activity and SFE at $z<3$. This picture should be taken just as an illustrative example, as 
the redshift-dependent mass scale at which halos can accrete intergalactic gas depends on the reionization redshift, the amplitude of the ionizing background,   
and the formation history of a halo rather than its instantaneous mass \citep{Noh14}. Yet, although the small number of isolated dwarfs with deep CMDs and 
concerns about the absolute 
ages of the earliest epochs of star formation in CMD-based SFHs urge caution in over-interpreting our findings, it is tempting to conjecture that it is 
the time-dependent suppression of baryonic infall by the UV background -- aided by stellar feedback -- that may regulate the fueling of dwarfs and 
drive the late evolution of their SFEs \citep[see also][]{Benitez14}. Regardless, there appears to be little need on these mass scales for additional feedback suppression at 
high redshifts -- dwarf galaxies were already in place at early times.

\acknowledgments
Support for this work was provided by NSF grant OIA-112445329745 and by NASA grant NNX12AF87G to PM, and by NASA through Hubble Fellowship grant 
HST-HF-51331.01, awarded by the STScI, to DRW.  We thank Peter Behroozi, Takashi Okamoto, and Sijing Shen for useful discussions
and assistance with Figures 2, 3, and 5.

\end{document}